# Regelation: why does ice melt under pressure?

*Phase-boundary reversible dispersivity and hydrogen-bond extraordinary recoverability*

Chang Q Sun

Ecqsun@ntu.edu.sg

Nanyang Technological University, Singapore

**Unlike other unusual materials whose bonds contract under compression, the O:H nonbond undergoes contraction and the H-O bond elongation towards O:H and H-O length symmetry in water and ice. The energy drop of the H-O bond dictates the melting point $T_m$ depression of ice. Once the pressure is relieved, the O:H-O bond fully recovers its initial state, resulting in Regelation.**

Ref:

[1] *Anomaly 2: Floating ice*, http://arxiv.org/abs/1501.04171

[2] *Anomaly 1: Mpemba effect*, http://arxiv.org/abs/1501.00765

[3] *Hydrogen-bond relaxation dynamics: resolving mysteries of water ice.* Coord. Chem. Rev., 2015. **285**: 109-165.

1  Anomaly: Ice Regelation

Observations in Figure 1 revealed the following:

1) Ice melts under pressure and freezes again when the pressure is relieved [1-4]**Error! Bookmark not defined.**. An ice block remains a solid after a weighted wire cutting it through [5].

2) Ice melts at a limit temperature of -22°C under 210 MPa pressure but a -95 MPa pressure (tension) raises the melting point up to +6.5°C [6, 7].

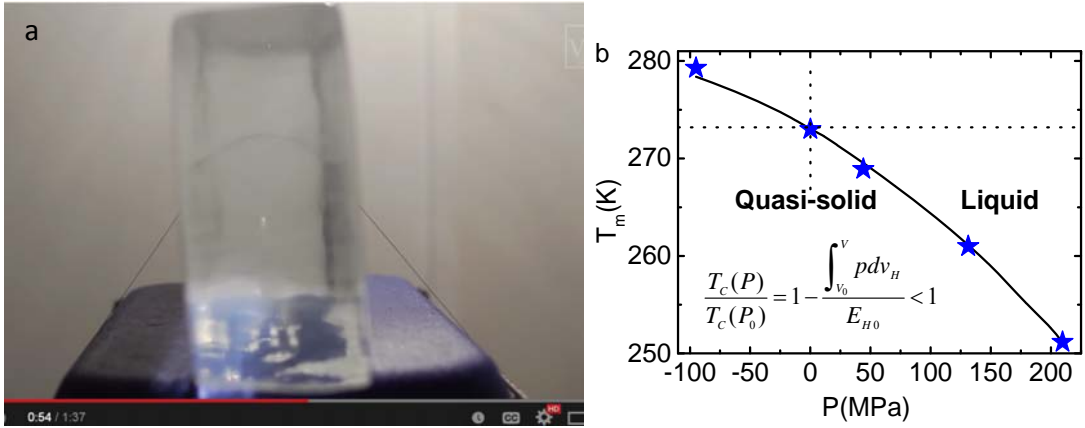

Figure 1  Regelation of ice. (a) A weighted wire cuts a block of ice through without severing it [5]. (b) Theoretical formulation [8] of the pressure dependence of the ice melting temperature $T_m(P)$ or the phase boundary between the liquid and quasi-solid [6, 7] indicates that the H-O bond energy relaxation dictates the $T_m(P)$.

2      *Reasons: hydrogen-bond memory and phase boundary dispersivity*

Figure 2 illustrates the following mechanisms:

1) A superposition of the segmental specific-heat $\eta_x(T/\Theta_{Dx})$ in Debye approximation of the O:H-O bond yields two intersecting temperatures that correspond to boundaries of the quasi-solid phase and the upper boundary closes to the melting point $T_m$ of ice (Figure 2a) [9].
2) The Debye temperature $\Theta_{Dx}$ correlates to $\omega_x$ in the $\hbar\omega_x \cong k\Theta_{Dx}$ relationship. Hence, any stimulation that relaxes the $\omega_x$ disperses the phase boundaries, accordingly [8].
3) Compression shortens the O:H nobond and raises its characteristic phonon frequency ($\omega_L$) but does the opposite to the H-O bond, which closes up the phase boundaries. Extension does the opposite. Quantitatively, the H-O bond energy dictates the $T_m$ as the inset equation in Figure 1b illustrates [10].
4) Oxygen atom always seeks for partners to retain its hybridized $sp^3$-orbit (Figure 2b) once the O:H nonbond breaks [11], which entitles O:H-O bond to recover completely from its relaxed or dissociated states once the excitation is relieved [12].

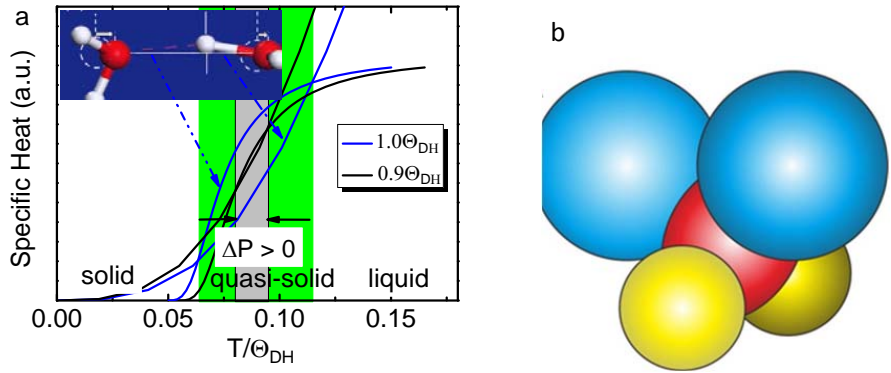

Figure 2. Schematic illustration of pressure squeezed quasi-solid phase boundaries and an $H_2O$ molecule. (a) An superposition of the specific-heat $\eta_x(T)$ curves for the O:H (x = L) and the H-O (x = H) segment defines two crossing points as boundaries of the quasi-solid phase. The high-temperature boundary corresponds to the temperature of maximal density (277 K) closing to the melting point $T_m$(273 K for bulk) and the lower to freezing ($T_N$ for homogeneous ice nucleation, 258 K for bulk). Compression ($\Delta P > 0$)/tension ($\Delta P < 0$) disperses the boundaries inwardly/outwardly by relaxing the O:H-O bond (inset) and the segmental $\omega_x$ that determines the $\Theta_{Dx}$. Excessive pressure (> 210 MPa) eliminates the phase boundaries, preventing $T_m$ from being further depressed. (b) Oxygen forms the tetrahedral structure of a water molecule at temperatures above 5 K [13] with two bonded $H^{+\delta}$ ions (yellow) and two $H^p$ dipoles (blue) induced by the lone pairs of the central $O^{2\delta-}$ ion (red; $\delta < 1$ is the charge quantity, which is assumed unity for discussion convenience)[12].

## 3    *Examples of Regelation*

The regelation is exceedingly interesting, because of its relation to glacial action under nature circumstances [14], in its bearing upon molecular action [15], and damage recovery of living cells. Observations indicate that O:H-O bond has extraordinary ability of recovering its relaxation and damage [16].

### 3.1    Ice cutting

One example of regelation can be seen by looping a fine wire over an ice cube and attaching a heavy weight to the wire (Figure 1a). The pressure of the wire exerts on the ice will melt the local ice gradually until the wire passing through the entire block of ice. The water refreezes behind the path of the wire so one can pull the wire through the ice, while leaving the ice cube intact. While regelation is occurring, some of the melting may be caused by heating conduction of the wire under tension. A molecular-dynamics (MD) study of a nanowire cutting through ice suggests that the transition mode and the cutting rate depend on the wetting properties of the wire - hydrophobic and thicker wires cut ice faster [17]. A video clip shows that a copper wire cutting ice faster than a fishing wire because of thermal conductivity [18].

### 3.2 Glacier: Source of river

Regelation occurs in glaciers, which forms the sources of Rivers like Yangze. Glaciers of Himalayas Mountains (8881 km apex in the boundary of India and Chania) in Asia source the India River, Yaluzangbu River, and Heng River. The mass of a glacier allows it to exert a sufficient amount of pressure on its lower surface to lower the melting point of the ice at its base, melting the ice and allowing the glacier to slide over the liquid. Under the right conditions, liquid water can flow from the base of a glacier to lower altitudes when the temperature of the air is above the freezing point of water [1, 2] (left panel of Figure 3).

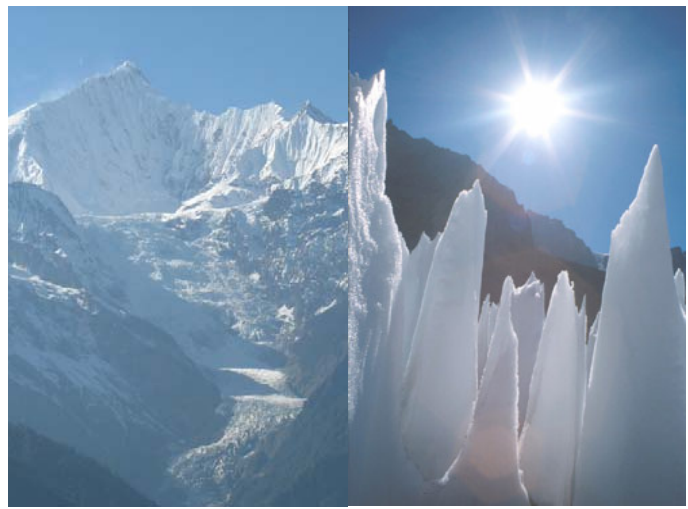

Figure 3 Glacier regelation and spiky ice. (Left) Meili Mountain in Yuannan, China, located at the junction of Nu river, Lanchang river, and Jinsha river (Photograph by Yi Sun, 2010). (Right) Spiked ice (called penitentes that can be 4 meters high) formed naturally on high-altitude glaciers in the relative humidity of ~70% and the -10 and -20 °C temperature range under irradiation. (Reprinted with permission from [19, 20].)

### 3.3  Spiky ice

The spikes of snow or ice are called penitentes, and some can be 4 meters high. They are common on high-altitude glaciers, such as those in the Andes Mountains, where the air is dry, and the sun's rays can turn ice directly into water vapor without melting it first–a process called sublimation. An initially smooth snow surface first develops depressions as some regions randomly sublimate faster than others. The curved surfaces then concentrate sunlight and speed up sublimation in the depressions, leaving the higher points behind as forests of towering spikes. At the micro-scale, similar-looking spikes help solar cell surfaces maximize their sunlight absorption. Penitentes grow faster at temperature between -10 and -20 °C under flood lamp irradiation and RH 70% coarsen [20].

According to the current premise, Spike ice formation requires suitable extrinsic conditions of temperature, pressure, humidity, and irradiation and intrinsic response of the O:H-O bond to these stimuli. Both the lower pressure at higher altitude and the undercoordination of molecules raises the $T_m$ of ice at apex so ice melts first at the dents, resulting in the spiked ice, as shown in Figure 3 (right).

### 3.4 Freezing, melting, and boiling

The H-O bond relaxation determines the melting point and the O:H freezing and boiling. Figure 4 shows that both air pressure and water boiling point drops as one goes higher. The lowered pressure lengthens and softens the O:H nonbond and the H-O bond relaxes reversely. The former determines the critical temperature of noiling and freezing and the latter freezing.  It is known that the boiling point becomes lower as one move to high altitude. At 3000 m height, the air pressure is 70 % and the boiling point is 87 °C. At 20000 m height (of a normal airplane flight) the pressure is almost zero and the boiling point will be 40 °C. The airplane must be pumped in air to maintain the cabinet pressure. The melting point, however, increases slightly only by 0.0072 °C at 20000 m height. The freezing point will drop a little as goes to 2000 m high.

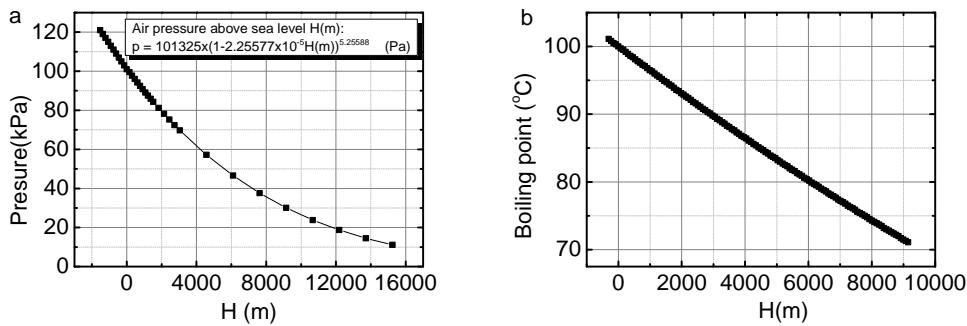

Figure 4 (a) Pressure and (b) boiling point as a function of altitude height. (Data sourced from [21, 22].)

*4 Notes on existing mechanisms*

### 4.1     Proton centralization in Phase X

In 1972, Holzapfel [23] firstly predicted that, under compression, an O:H-O bond might be transformed from the highly asymmetrical O:H-O configuration to a symmetrical state in which the H proton lies midway between two $O^{2-}$ ions, leading to a non-molecular symmetrical phase of ice-X. Goncharov et al. [24] confirmed this prediction in 1998 using *in situ* high-Pressure Raman spectroscopy. The proton centralization in the O:H-O bond of ice-VIII occurred at about 60 GPa and 100 K, and no further shift of phonon frequency was observed when they increased the pressure, since the O:H and H-O had both reached identical lengths (0.11 nm) [25, 26]. Proton centralization also occurs in liquid $H_2O$ at 60 GPa and 85 K, and to liquid $D_2O$ at 70 GPa and 300 K [27].

Compression-induced proton centralization evolves the pairing potential wells into a single well located midway between $O^{2-}$ ions [28] which was attributed to "translational proton quantum tunneling" [25, 29, 30] or to the extraordinary, yet unclear, behavior of the inter- and intramolecular bonds [31].

### 4.2     Temperature dependence of compressibility

Compression shortens the O-O distance but lengthens the H-O bond, resulting in the low compressibility of ice compared to 'normal' materials [32]. The compressibility of liquid water is slightly higher than that of ice. If a liquid water is sufficiently cold, its diffusivity increases and its viscosity decreases upon compression. In contrast, compression of most other liquids leads to a progressive loss of fluidity as molecules are squeezed closer together [33].

The compressibility of water is a function of pressure and temperature. As shown in Figure 5, at 0 °C and zero pressure, the compressibility is $5.1\times10^{-10}$ $Pa^{-1}$. At the zero-pressure limit, the compressibility reaches a minimum of $4.4\times10^{-10}$ $Pa^{-1}$ around 45 °C before increasing again with increasing temperature. As the pressure is increased, the compressibility decreases, being $3.9\times10^{-10}$ $Pa^{-1}$ at 0 °C and 100 MPa [34].

The bulk modulus of water is 2.2 GPa. The low compressibility of water, leads to its often being assumed as incompressible. The low compressibility of water means that even in the deep oceans at 4 km depth, where pressures are 40 MPa, there is only a 1.8% decrease in volume.

Heating depolarization of water (inset in Figure 5) takes the responsibility for compressibility incensement at high temperature [35]. The inset shows the heating and salting effect on the contact angle between a deionized water droplet and a glass substrate. Salting enhances but heating demotes the polarization that provides repulsive force between oxygen ions. For instance, the contact angle drops from 47 to 30 ° when the droplet is heated from 20 to 80 °C. The weakening of the inter ions repulsion results in the recovery of the compressibility.

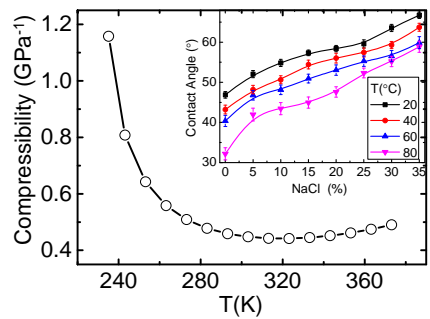

Figure 5. Compressibility of water and ice as a function of temperature shows the anomaly of extremely at 320 K. [36]

## 4.3 Ice skating

Ice skating is given as an example of regelation to create liquid to lubricate ice; however the pressure required is much greater than the weight of a skater. Additionally, regelation does not explain how one can ice skate at temperatures below the limit of -22 °C.

If the contacting area of the skate to ice is 150×10$^{-6}$ m$^2$ (1 mm wide and 150 mm long) and the skater weighs 500 Newtons, the pressure applied will be 3.3 MPa. As the melting point of ice falls by 0.0072 °C for each additional atm (0.1 MPa) of pressure applied, the melting point will drop by 0.24 °C only. Therefore, skating provides insufficient pressure for melting ice (Figure 6).

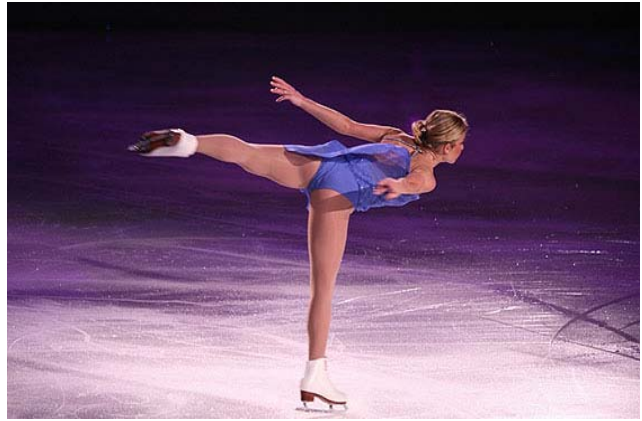

Figure 6. Ice skating provides insufficient pressure to melt ice but lowers the melting point by ¼ °C. (Galina Barskaya/Dreamstime.com).

## 4.3 Negative thermal expansion

It might be true that regelation can occur for substances with the property of expanding upon freezing, but mechanisms of the freezing expansion and for the regelation remain yet unclear [9]. Faraday concluded that this phenomenon occurs only to ice after conducting numerous experiments in 1859 [4] "Many salts were tried (without much or any expectation),-crystals of them being brought to bear against each other by torsional force, in their saturated solutions at common temperatures. In this way the following bodies were experimented with: -Nitrates of lead, potassa, soda; sulphates of soda, magnesia, copper, Zinc; alum; borax; chloride of ammonium; ferro-prussiate of potassa; carbonlate of soda; acetate of lead; and tartrate of potassa and soda; but the results with all were negative. My present conclusion therefore is that the property is special for water; and that the view I have taken of its physical cause does not appear to be less likely now than at the beginning of this short investigation, and therefore has not sunk in value among the three explanations given."

## 5 Historical background

Discovered by Faraday [4], Thomson [37], and Forbes [1, 2] in 1850's (Figure 7), regelation is the phenomenon of ice melting under pressure and freezing again when the pressure is relieved at temperatures around -10 °C. Faraday noted in his paper that [4], 'Two pieces of thawing ice, if put together, adhere and become one; at a place where liquefaction was proceeding, congelation suddenly occurs. The effect will take place in air, in water, or *in vacuo*. It will occur at every point where the two pieces of ice touch; but not with ice below the freezing point, i.e., with dry ice, or ice so cold as to be everywhere in the solid state'. The generally accepted explanation for the phenomena is that sufficient compressive stresses exist at the contact area to cause melting when the pieces of ice are brought together, and when this stress is released solidification occurs. This was originally proposed by James Thomson [3] and endorsed by his brother, Lord Kelvin (William Thomson) (Figure 8).

James Thomson noted [37]: "The phenomenon (Regelation) is a consequence of the properties, announced from theory by Prof. James Thomson, and then exemplified by an experiment; and the explanation depends on the theories put forward by him-the first (1857) founded on the lowering of the freezing point of water by pressure, and the second (1861) founded on the tendency to melt given by the application to the solid ice of forces whose nature is to produce change of form as distinguished from forces applied alike to the liquid and solid. The stress upon the ice, due to its pressure on the network, gives it a tendency to melt at the point in contact with the wire, and the ice, in the form of water intermixed with fragments and new crystals, moves so as to relieve itself of pressure."

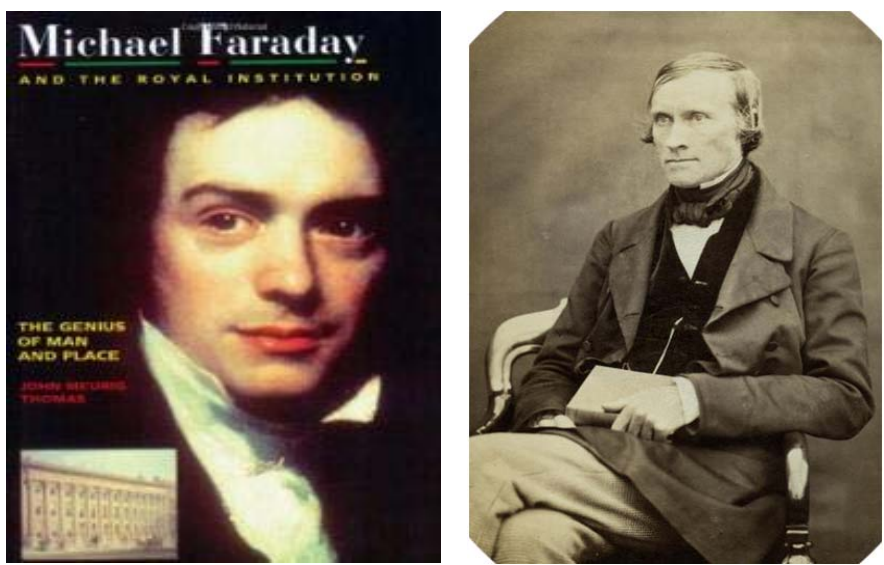

Figure 7 (left) Michael Faraday [38] FRS (22 September 1791 – 25 August 1867) was an English scientist who contributed to the fields of electromagnetism and electrochemistry. (Right) James

David Forbes FRS FRSE FGS (20 April 1809 – 31 December 1868) was a Scottish physicist and glaciologist who worked extensively on the conduction of heat and seismology.

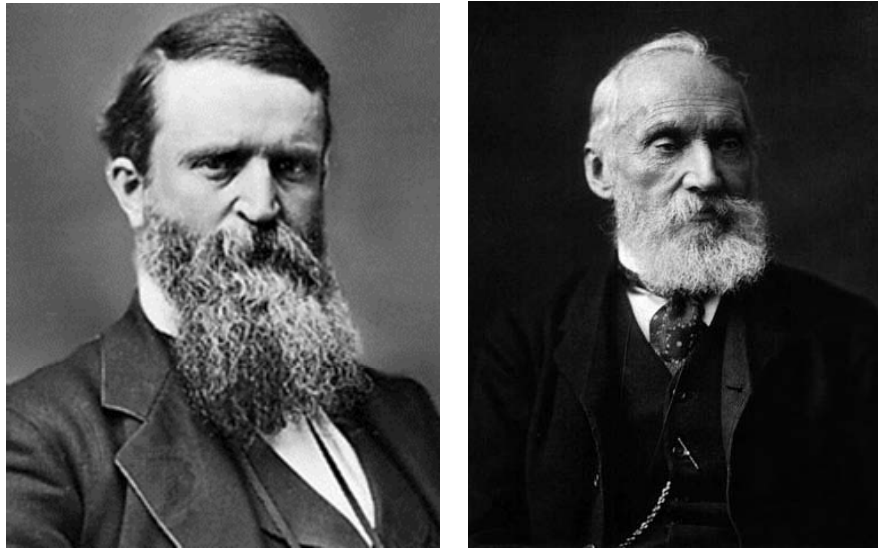

Figure 8 (Left) **James Thomson** (16 February 1822 – 8 May 1892) was an engineer and physicist whose reputation is substantial though it is overshadowed by that of his younger brother (Right) William Thomson (Lord Kelvin). (26 June 1824 – 17 December 1907). Lord Kevin was a British mathematical physicist and engineer. At the University of Glasgow he did important work in the mathematical analysis of electricity and formulation of the first and second laws of thermodynamics, and did much to unify the emerging discipline of physics in its modern form. For his work on the transatlantic telegraph project he was knighted by Queen Victoria, becoming Sir William Thomson. He had extensive maritime interests and was most noted for his work on the mariner's compass. (Free Wikimedia)

However, consistent understanding or numerical reproduction of regelation has yet been achieved since 1850's when Faraday, Thomson, and Forbes debated on possible mechanisms from the classical thermal dynamics and quasi-liquid skin viewpoints [4, 39]. Faraday supposed that a particle of water, which could retain the liquid state whilst touching ice only on one side, could not retain the liquid state if it were touched by ice on both sides; but became solid, the general temperature remaining the same. Thomson who discovered that pressure lowered the freezing-point of water, attributed the regelation to the fact that two pieces of ice could not be made to bear on each other without pressure; and that the pressure, however slight, would cause fusion at the place where the particles touched, accompanied by relief of the pressure and resolidification of the water at the place of contact. Forbes assented to neither of these views; but due to the gradual liquefaction of ice, and assuming that ice is essentially colder than ice-cold water, i.

e. the water in contact with it, he concluded that two wet pieces of ice will have the water between them frozen at the place where they come into contact.

6      Quantitative evidence

6.1    O:H-O bond length symmetrization and compressibility

Generally, compression shortens all bonds of a normal substance. However, for the O:H-O bond in water and ice, compression shortens the softer O:H nonbond and lengthens the stiffer H-O by different amounts. The O:H shortens more than the H-O elongation through inter oxygen ions repulsion. Figure 9a shows the $V/V_0(P)$ profiles of water (300 K) and ice (77 K) measured using *in situ* high-pressure and low-temperature synchrotron XRD and Raman spectroscopy [40]. Molecular dynamics (MD) calculations converted the $V/V_0$–$P$ profiles into the $d_x/d_{x0} - P$ curves [10]. As shown in Figure 9b, compression shortens the O:H nonbond from .1.767 to 0.1.692 Å, meanwhile lengthens the H-O bond from 0.974 to 1.003 Å when the pressure is increased from 1 to 20 GPa [41-43]. An extrapolation of the $d_x/d_{x0} - P$ curves meet at the exact point of proton centralization occurred in phase X [24, 25, 40] at 58.6-59.0 GPa with an O-O distance of 2.21-.2.20 Å [25].

Therefore, the proton centralization arises from pressure-induced O:H-O bond asymmetrical relaxation rather than from transitional H$^+$ proton quantum tunneling that is unlikely because of the strong H-O bond of 3.97 eV energy. Constrained by measured proton centralization and the $V/V_0$–$P$ profiles, the $d_x/d_{x0} - P$ curves represent the true situation in large temperature range (77-300 K) irrespective of the probing techniques or conditions.

The following polynomials formulate the O:H-O bond length relaxations, in which $P_0 = 0.1$ MPa is the atmospheric pressure that approximates to zero:

$$\begin{pmatrix} d_H / 0.975 \\ d_L / 1.768 \\ V / 1.060 \end{pmatrix} = 1 + \begin{pmatrix} 9.510 & 2.893 \\ -3.477 & -10.280 \\ -238.000 & 47.000 \end{pmatrix} \begin{pmatrix} 10^{-4} P^1 \\ 10^{-5} P^2 \end{pmatrix}.$$

(1)

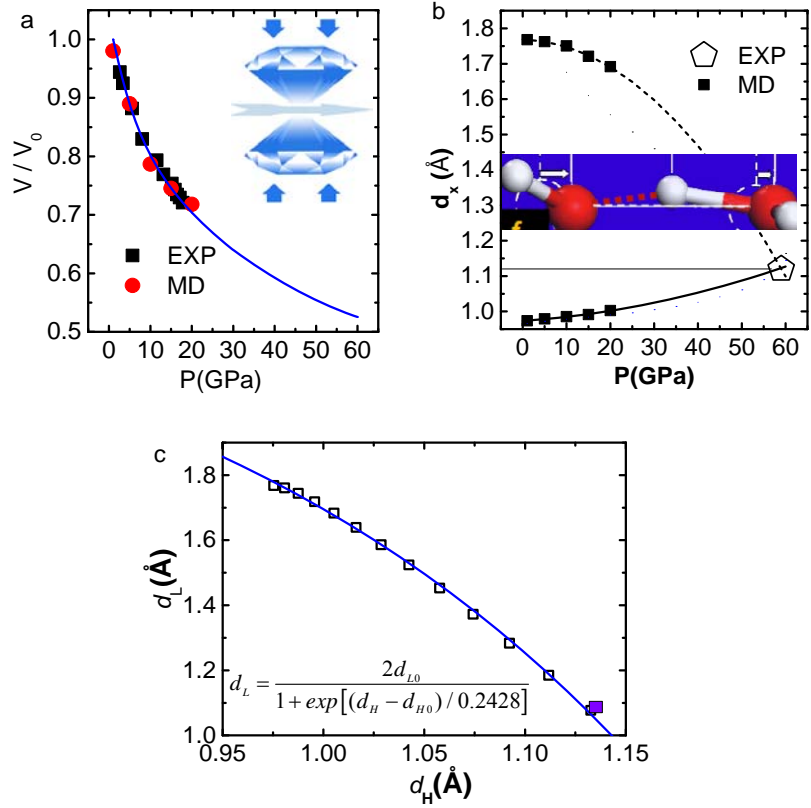

Figure 9. MD conversion of the measured (a) $V/V_0$–P profiles [40] (inset illustrates experimental set up [44]) into (b) the $d_x$–P curves meeting at the point of proton centralization occurring at 59 GPa and 2.20 Å [24, 25]. Hiding the variable P results in (c) the $d_L$-$d_H$ correlation being independent of experimental conditions or probing method in the temperature range of 77-300 K. (Reprinted with permission from [10].)

Taking the $P$ as a hidden parameter, the $d_x/d_{x0} - P$ profiles give the universal $d_L$–$d_H$ cooperativity, shown in Figure 9c, for the mean size, separation, and density of molecules packed in water and ice in the tetrahedrally coordinated yet fluctuating structure [45]:

$$\begin{cases} d_{oo} = d_L + d_H = 2.6950\rho^{-1/3} & (Molecular\ separation) \\ d_L = \dfrac{2d_{L0}}{1+exp\left[(d_H - d_{H0})/0.2428\right]}; & (d_{h0}=1.0004, d_{L0}=1.6946) \end{cases}$$

(2)

Thus, given any known value of the $d_H$, $d_L$, $d_{O-O}$ and $\rho$ of $H_2O$, one can determine all the rest provided with involvement of the inter-oxygen Coulomb repulsion. Molecules in the low-density vapor phase may not be subject to this formulism because of the extremely weak O:H interaction. Table 1 features the DFT-MD derivatives of the O:H-O bond segmental relaxation dynamics and the band gap change as a function of pressure. Discrepancy in numerical derivatives exists so one needs to focus on the trend of change instead of the accuracy.

Table 1. Pressure dependence of the mass density $\rho$, segmental lengths $d_x$, O-O distance $d_{O-O}$, and band gap $E_G$ of ice [10].

| | DFT | | | | MD | | |
|---|---|---|---|---|---|---|---|
| $P$(GPa) | $\rho$ (g/cm$^3$) | $d_H$ (Å) | $d_L$ (Å) | $E_G$ (eV) | $d_H$ (Å) | $d_L$ (Å) | $d_{O-O}$ (Å) |
| 1 | 1.659 | 0.966 | 1.897 | 4.531 | 0.974 | 1.767 | 2.741 |
| 5 | 1.886 | 0.972 | 1.768 | 4.819 | 0.979 | 1.763 | 2.742 |
| 10 | 2.080 | 0.978 | 1.676 | 5.097 | 0.985 | 1.750 | 2.736 |
| 15 | 2.231 | 0.984 | 1.610 | 5.353 | 0.991 | 1.721 | 2.713 |
| 20 | 2.360 | 0.990 | 1.556 | 5.572 | 1.003 | 1.692 | 2.694 |
| 25 | 2.479 | 0.996 | 1.507 | 5.778 | – | – | – |
| 30 | 2.596 | 1.005 | 1.460 | 5.981 | | | |
| 35 | 2.699 | 1.014 | 1.419 | 6.157 | | | |
| 40 | 2.801 | 1.026 | 1.377 | 6.276 | | | |
| 45 | 2.900 | 1.041 | 1.334 | 6.375 | | | |
| 50 | 2.995 | 1.061 | 1.289 | 6.459 | | | |
| 55 | 3.084 | 1.090 | 1.237 | 6.524 | | | |
| 60 | 3.158 | 1.144 | 1.164 | 6.590 | | | |

6.2     Cooperative phonon relaxation: Coulomb coupling

Generally, compression stiffens all phonons of 'normal' substance such as carbon allotropes [46], ZnO [47], group IV [48], group III-V [49], and group II-VI [50] compounds without exception; however, for ice and water, compression stiffens the softer O:H stretching phonons ($\omega_L$ < 300 cm$^{-1}$) but softens the stiffer H-O stretching phonons ($\omega_H$ > 3000 cm$^{-1}$) cooperatively [40, 51-54].

Figure 10 shows the compression-induced $\omega_x$ cooperative relaxation of ice-VIII phase at 80 K [6, 103] and the $\omega_H$ for liquid water at 296 K up to 0.4 GPa pressure [54]. DFT and quantum Monte Carlo calculations suggest that the O:H interaction contributes positively, while the H-O bond negatively to the lattice energy as the pressure is increased [55]. Both phonon and energy of the segmented O:H-O bond relax consistently and cooperatively under compression.

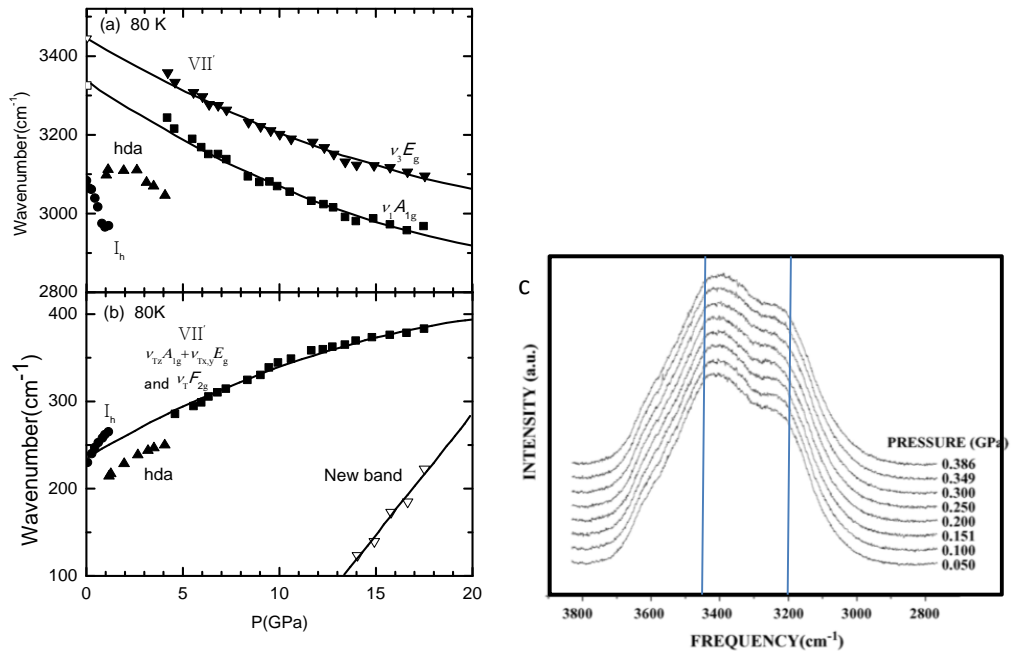

Figure 10. Compression (a) softens (redshift) the $\omega_H$ and (b) stiffens (blueshift) the $\omega_L$ cooperatively and monotonically for ice and softens the $\omega_H$ for water (3450 cm$^{-1}$ for skin and 3200 cm$^{-1}$ for bulk) at 296 K up to 0.4 GPa. (Reprinted with permission from [53, 54].)

Figure 11a features the MD-derived phonon relaxation as a function of pressure, which agree with trends probed using Raman and IR spectroscopy from ice-VIII at 80 K [40, 51, 52]. Compression softens the $\omega_H$ from 3520 cm$^{-1}$ to 3320 cm$^{-1}$ and stiffens the $\omega_L$ from 120 to 336 cm$^{-1}$, disregarding the possible phase change and other supplementary peaks nearby. Figure 11b compares the calculated and a collection of the measured $\Delta\omega_x$ for ice. Consistency in the pressure-derived $\omega_x$ cooperativity in Figure 11b for both water [54] and ice [51, 52, 56] confirms that compression shortens and stiffens the O:H nonbond and relaxes the H-O bond reversely in all the liquid and solid forms of water.

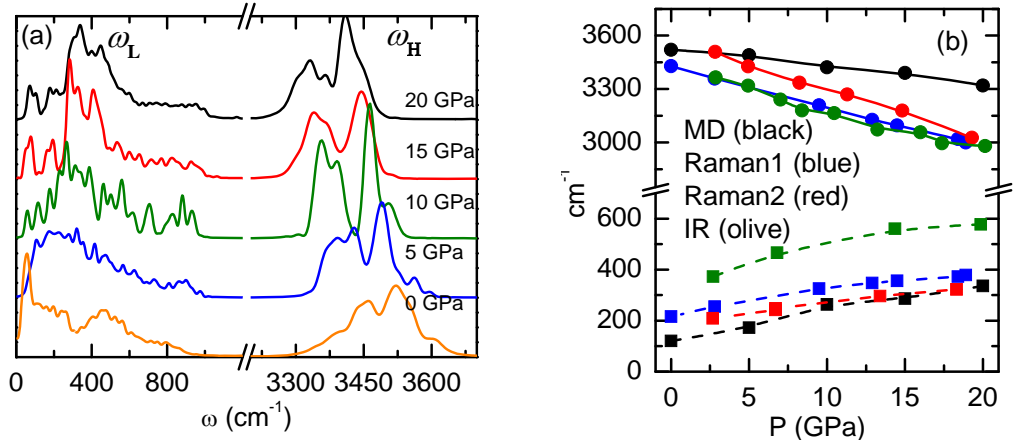

Figure 11. (a) MD-derived $\omega_x$ relaxation and (b) trend agreement between Raman and IR measurements and calculations for the ice-VIII phase at 80 K [40, 51, 52]. (Reprinted with permission from [10].)

6.3    O:H-O bond potentials: extraordinary recoverability

Figure 12a shows a pairing potential u(r) for the dimer bond in a regular substance. The coordinates ($d$, $E_b$) at equilibrium correspond to the bond length and bond energy, which relax under an external stimulus, regardless of the shape of the particular u(r). A Taylor series approximate the u(r) as follows:

$$u(r) = \left.\frac{\partial^n u(r)}{n!\partial r^n}\right|_{r=d} x^n = E_b + \left.\frac{\partial^2 u(r)}{2\partial r^2}\right|_{r=d} x^2 + \left.\frac{\partial^3 u(r)}{6\partial r^3}\right|_{r=d} x^3 + 0(x^{n\geq 4})$$

(3)

The zeroth differential is the bond energy at equilibrium $E_b$. Higher-order differentials corresponding to the harmonious and nonlinear vibrations determine the shape of the u(r). The vibration amplitude $x$ is 3% or less than atomic distance $d$ of the substance at temperatures below melting. The second derivative defines the harmonious vibration frequency of the dimer with reduce mass μ:

$$\frac{1}{2}\mu\omega^2 x^2 = \left.\frac{\partial^2 u(r)}{2\partial r^2}\right|_{r=d_x} \propto \frac{E_x}{d_x^2}$$

$$\Rightarrow \omega_x^2 \propto \frac{E_x}{\mu d_x^2}$$

(4)

Generally, external stimuli, such as stressing and heating modulate the length $d(T, P)$ and energy $E(T, P)$ of the representative bond along a path function $f(T, P)$ [57]. For instance, compression stores energy into a substance by shortening and stiffening all bonds with possible plastic deformation when the compression is relieved. Tension does the opposite [58]. The following formulates bond relaxation in length and energy under stimuli (P, T) [12]:

$$\begin{cases} d(P,T) = d_b \left(1 + \int_{T_0}^{T} \alpha(t) dt\right)\left(1 - \int_{P_0}^{P} \beta(p) dp\right) \\ E(P,T) = E_b \left(1 - \dfrac{\int_{T_0}^{T} \eta(t) dt + \int_{V_0}^{V} p(v) dv}{E_b}\right) \end{cases}$$

(5)

where $T_0$ and $P_0$ are the ambient referential conditions. The $\alpha(t)$ is the thermal expansion coefficient. $\beta = -\partial v/(v\partial p)$ is the compressibility ($p < 0$, compressive stress) or extensibility ($p > 0$ tensile stress). The $v$ is the volume of a bond (the product of length and its cross-sectional area). The $\eta(t)$ is the specific heat of the representative bond in Debye approximation. The Debye temperature $\Theta_D$ determines the rate of the $\eta(t)$ curve to it saturation. The integration of the $\eta(t)$ from 0 K to the $T_m$ equals the bond energy under constant pressure [8].

A Lagrangian-Laplace transformation of the measured $d_x$ and $\omega_x$ (Figure 9b and Figure 10a) for the O:H-O bond oscillating pair turns out the segmental force constant $k_x$ and energy $E_x$ (below 40 GPa), which maps the potential paths of the O:H-O bond relaxation under compression, see Figure 12b [59],

$$\begin{pmatrix} \omega_H / 3326.140 \\ \omega_L / 237.422 \\ E_H / 3.970 \\ E_L / 0.046 \end{pmatrix} = 1 + \begin{pmatrix} -0.905 & 1.438 \\ 5.288 & -9.672 \\ -1.784 & 3.124 \\ 25.789 & -49.206 \end{pmatrix} \begin{pmatrix} 10^{-2} P^1 \\ 10^{-4} P^2 \end{pmatrix}.$$

As shown in Table 2, compression increase the $E_L$ from 0.046 to 0.250 eV up to 40 GPa and then decrease to 0.16 eV; the $E_H$ decreases monotonically from 3.97 eV to 1.16 eV at 60 GPa. Different from 'normal' situation in Figure 12a, compression raises the total energy of the O:H-O bond rather than lowers it, as shown in Figure 12c, which defines the O:H-O bond to fully recover its initial states once the compression is relieved without any plastic deformation. Therefore, ice melts under compression and freezes again when the pressure is relieved.

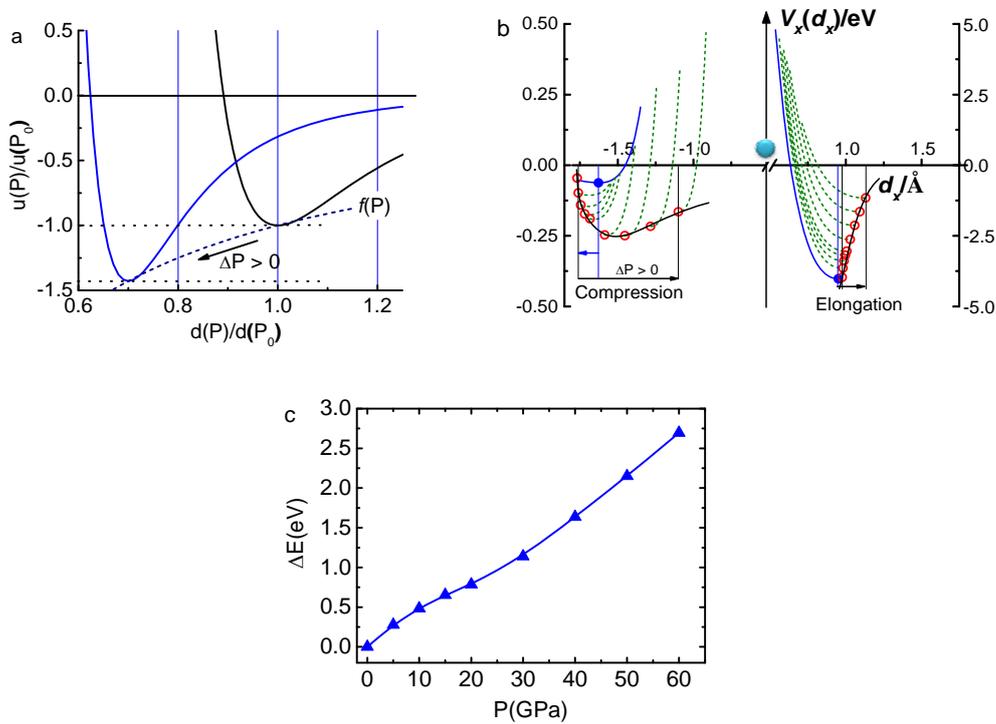

Figure 12. (a) The long-range, mono-well potential for paring atoms in a 'normal' substance [58] and (b) the asymmetrical, short-range, double-well potential paths for the O:H-O bond under compression (l. to r.: P = 0, 5, 10, 15, 20, 30, 40, 50, 60 GPa) [60, 61]. Compression stores energy by shortening and stiffening the bond whereas tension does the opposite, along an $f(P)$ path in (a). O:H-O potentials include the O:H nonbond van der Waals like (vdW-like) interaction (maximal at $E_L \sim 0.25$ eV, left-handed side), the H-O exchange interaction ($E_H \sim 4.0$ eV, right-handed side), and the Coulomb repulsion (C-repulsion) between electron pairs on oxygen ions. (c) Compression raises the O:H-O bond energy consistently, which provides force driving the Regelation of ice. (Reprinted with permission from [8].)

Table 2. Pressure-dependence of the O:H-O segmental cohesive energy $E_x$ and the net gain at each quasi-equilibrium state under compression. Unlike 'normal' substance that gains energy with possible plastic deformation under compression, O:H-O bond always losses energy and tends to recover from its higher energy state to lower initial state without any plastic deformation.

| $P$ (GPa) | $E_L$ (eV) | $E_H$ (eV) | $E_{H+L}(P)-E_{H+L}(0)$ |
|---|---|---|---|
| 0 | 0.046 | 3.97 | 0 |
| 5 | 0.098 | 3.64 | -0.278 |
| 10 | 0.141 | 3.39 | -0.485 |
| 15 | 0.173 | 3.19 | -0.653 |
| 20 | 0.190 | 3.04 | -0.786 |
| 30 | 0.247 | 2.63 | -1.139 |
| 40 | 0.250 | 2.13 | -1.636 |
| 50 | 0.216 | 1.65 | -2.15 |
| 60 | 0.160 | 1.16 | -2.696 |

6.4     Band gap modulation: Blue iceberg

Usually icebergs are white because they are made of compressed snow, which reflects all frequencies of visible light. However, if high pressures squeeze the flakes together, or sea water freezes, the gaps between the snowflakes disappear, taking with them the equi-reflecting surfaces [62].

Long wavelengths of light from the sun (reds and yellows) are absorbed when passing through the ice, whereas blue light is scattered. Some of the scattered light is reflected to us, producing the blue color we associate with pure water – although microorganisms or chemicals can sometimes add a greenish tint. The same thing happens when the water is frozen, provided all the air has been eliminated. This occurs a lot more often at the bottom of large, old blocks of ice than the top, and is usually hidden from view.  Figure 13 shows the phonon of an iceberg flipped over in Antarctica taken by Alex Cornell came across the aftermath of one such event, in comparison with the ordinary iceburg.

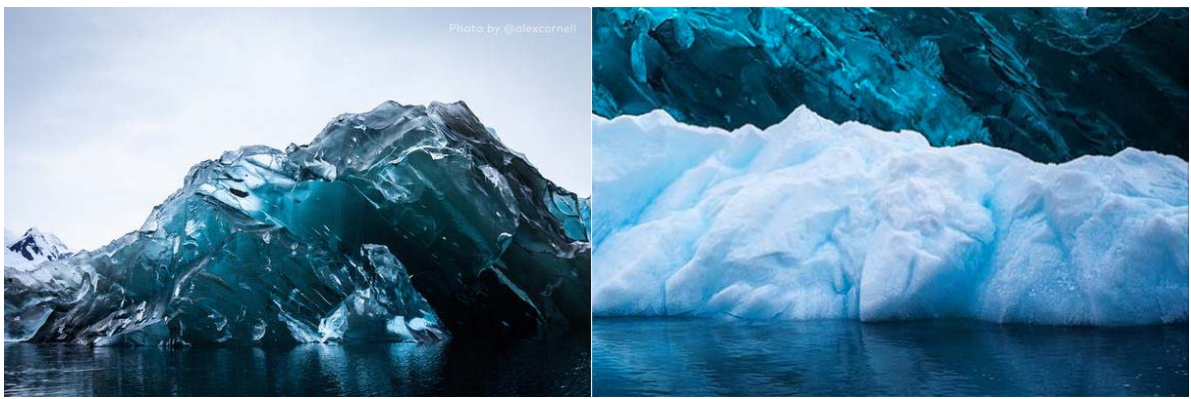

Figure 13 The underside (left) of an iceberg is strikingly deep blue in comparison with its upside that is white in Antarctica (photo credit: Alex Cornell [62]).

Figure 14 shows a collection of the UV absorption spectra by ice under different pressure [63]. The onset of UV absorption by ice, as an indication of band gap evolution, shifts positively with increasing pressure, making ice more transparent and blue (Figure 13). They attributed this effect to an increase of the Stark shift in water caused by the electrostatic environment at smaller volumes.

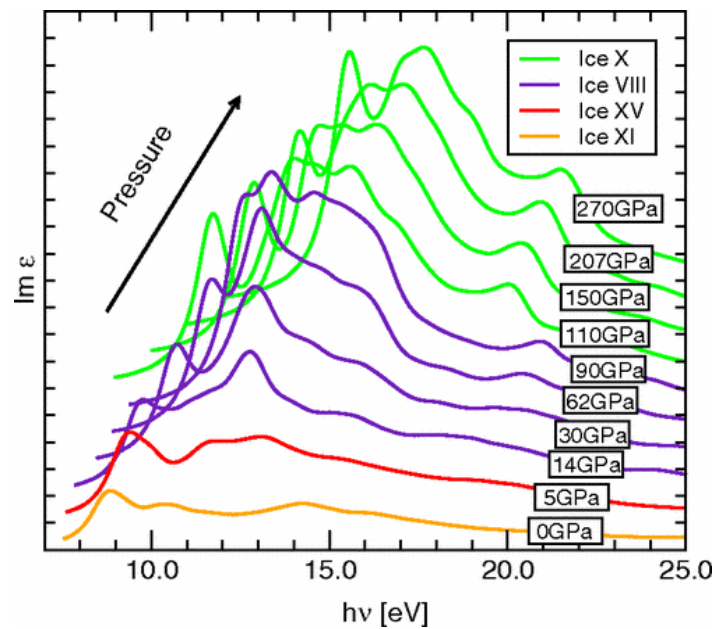

Figure 14 Many-body optical absorption spectra of various ice crystal structures under pressure. Spectra are offset vertically for clarity. (Reprinted with permission from [63].)

Generally, compression enlarges the optical band gap due to bond strength gain, since the band gap is proportional to the bond energy with involvement of electron–phonon coupling [47, 64]. Band gap expansion of ice follows the same pressure trend of "normal" substance but due to a different mechanism, because compression softens the H-O bond.

Figure 15a and Table 2 feature the DFT-derived DOS evolution of ice-VIII with pressure varying from 1 to 60 GPa. The bottom edge of the valence band shifts down from -6.7 eV at 1 GPa to -9.2 eV at 60 GPa, but its upper edge at the Fermi level remains unchanged. The conduction band shifts up from 5.0 to 12.7 eV at 1 GPa to 7.4–15.0 eV at 60 GPa. The band gap expands further at higher pressures, from 4.5 to 6.6 eV, as shown in Figure 15b, when the pressure is increased from 1 to 60 GPa.

The band gap expansion in compressed ice arises not from the $E_H$ but is caused by a different mechanism. The energy shift of the DOS above $E_F$ results from polarization of the lone pair by the entrapped core electrons. The energy shift of the valence DOS below $E_F$ arises from entrapment of the bonding states of oxygen [11, 65].

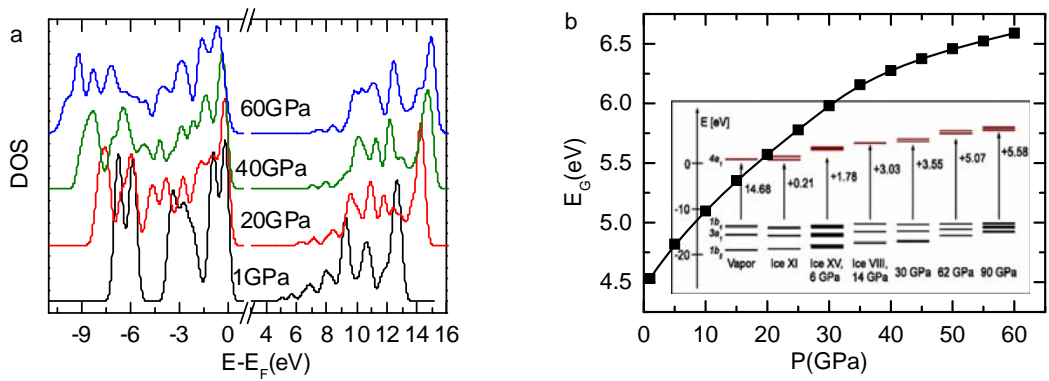

Figure 15. Compression widens the band gap of ice-VIII. (a) Compression entraps the valence DOS (bottom edge) and polarizes the conduction DOS, resulting in (b) band gap expansion. The upper edge of the valence band in (a) at the $E_F$ is conserved. Inset b shows the occupied (black lines) and unoccupied (red lines) energy levels of water in vapor and molecular crystal structures, with a single-particle gap increase relative to vapor phase [63]. (Reprinted with permission from [10].)

6.5     Regelation: O:H-O memory and phase boundary dispersivity

The following clarifies the remaining questions regarding Regelation:

1) Why the $E_H$ instead of the $E_L$ dominates the $T_C$ and $T_m$?
2) Why does ice at the lowest limit of -22 °C?
3) Why the molten ice freezes again when the pressure is relieved?

As shown in Figure 2a, the vibration frequency $\omega_x$ determines the Debye temperature and the cohesive energy $E_x$ determines integral of the respective segmental specific heat curve $\eta_x$. The supposition of these two $\eta_x$ curves defines two intersecting points that correspond to the phase boundaries of the quasi-solid phase or the extreme density temperatures. The high-temperature boundary is sensitive to the $\eta_H$ rather than the $\eta_L$, and therefore, the $E_H$ and $\omega_H$ determine the $T_m$.

Compression/tension ($\Delta P > 0$)/($\Delta P < 0$) disperses these boundaries simultaneously and reversely by modulating the $\Theta_{Dx} \propto \omega_{Dx}$, on the base of an extension of Einstein's relationship [8]: $\Theta_{DL}/\Theta_{DH} \approx \omega_L/\omega_H$. As shown in Figure 10, compression stiffens the $\omega_L$ and meanwhile softens the $\omega_H$, which closes up the crossing temperatures and lowers the $T_m$. Negative pressure does the opposite to raise the $T_m$. The entire process is reversible according to total energy change, shown Figure 12c.

The $E_L$ determines molecular dissociation turning liquid water into vapour at $T_V$. If remove one $H_2O$ molecule from the bulk, one has to break four O:H nonbonds with energy of 0.38 eV per molecule at the ambient pressure [66]. As elaborated above, the $T_m$ depends on the $E_H$ though the $T_m$ is lower than the $T_V$ at the ambient pressure.

Regelation happens under the conditions of -95 MPa $\leq$ P $\leq$ 210 MPa and 252 K $\leq$ T < 273 K. If the pressure is higher than the critical value of 210 MPa, the quasi-solid phase disappears and the liquid transits into phase III, VI and V.

Once the O:H nonbond breaks by perturbation, oxygen atoms will find new partners to retain the $sp^3$-orbital hybridization, which is the same to diamond oxidation and metal diffusive corrosion [11]. Therefore, O:H-O bond has the strong recoverability for relaxation and dissociation without any plastic deformation. Compression stores energy into water ice by raising the total bond energy, $E_L+E_H$, through inter electron pair repulsion [16], once the pressure is relieved, the O:H-O bond will relax to its initially low energy states.

7    Further extension
7.1    Boundary formulation of water phase diagram

The phase diagram of water and ice in Figure 16 shows the following boundary $T_C(P)$ features according to their slopes:

$$\frac{dT_C}{dP} = \begin{cases} \cong \infty & (X-XI, X-(VII,VIII), etc) \\ \cong 0 & (Ic-VI, XV-VI, etc) \\ > 0 & (Liquid-Vapour, Liquid-(III,IV,V,VII)) \\ < 0 & (VII-VIII, Ih-Liquid, etc) \end{cases}$$

(6)

The versatile $T_C(P)$ slopes represent different yet unclear dynamics of O:H-O bond relaxation at the phase boundaries. Numerous formulae described the $T_C(P)$ from the classical thermodynamic point of view, mainly for the Liquid-Vapor phase transition. For instances, Clausius–Clapeyron equation [67] describes water vapor under typical atmospheric conditions (near standard temperature and pressure) and August-Roche-Magnus formula [68] approximates the temperature dependence of the saturation vapor pressure $P_S$:

$$\begin{cases} \dfrac{dP_s}{dT} = \dfrac{L_v(T)P_s}{R_v T^2} & (Clausius-Clapeyron) \\ P_s(T) = 6.1094 \exp\left(\dfrac{17.625T}{T+243.04}\right) & (August-Roche-Magnus) \end{cases}$$

(7)

Where $L_v$ is the specific latent heat of evaporation of water and $R_S$ is the gas constant of vapor.

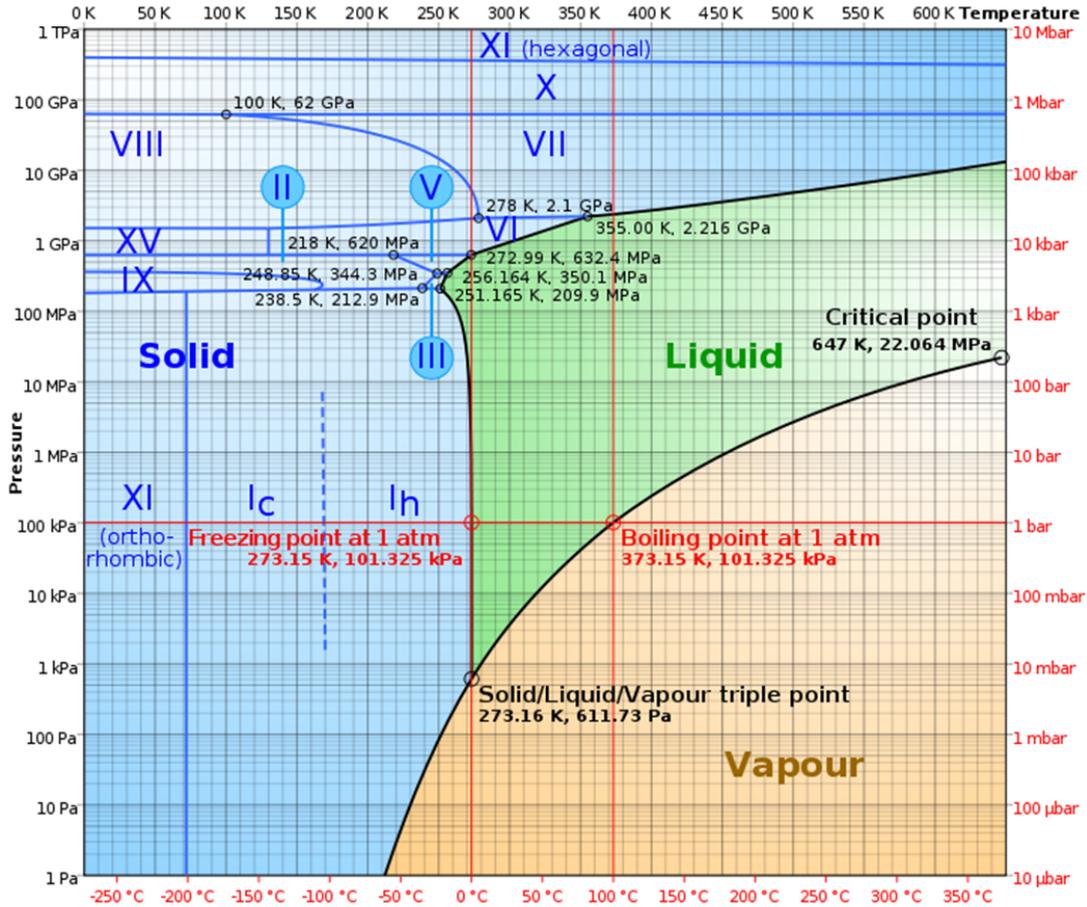

Figure 16. Phase diagram for water [69]. $dT_C/dP < 0$ at boundaries for II-V, Liquid-Ih, and VII-VIII transition indicates that H-O bond elongation dominates. $dT_C/dP > 0$ at boundaries for Liquid-Vapor, Liquid-(V, VI, VII) transition suggests that O:H nonbond contraction governs the $T_C(P)$ elevation. $dT_C/dP \cong \infty$ at the (VII, VIII)→X boundary occurring at 60 GPa corresponds to the O:H and H-O length symmetrization. Compression promotes yet heating demotes the proton centralization [70-72]. $dT_C/dP \cong 0$ at Ih-XI boundary in low temperatures ($\eta_x \cong 0$) indicates bond angle relaxation dominance with negligible change in the segmental lengths.(Reprinted with permission from [69, 71].)

7.2  H-O bond elongation dominance: $dT_C/dP < 0$

For other 'normal' substance, $T_C$ is proportional to the atomic cohesive energy, $T_C \propto zE_z$, where $z$ is the effective atomic CN and $E_z$ is the cohesive energy of bond between the $z$-coordinated atoms [73]. However, for water molecules, the $T_C$ is proportional to $E_x$ where the x needs to be certain. Molecular CN contributes indirectly to the bond length and energy but not to the $T_C \propto E_x$, because of the 'isolation' of the $H_2O$ molecule by its surrounding lone pairs.

It is universally true that compression shortens the softer O:H nonbond more than the stiffer H-O bond elongates because of the repulsion between electron pairs on adjacent O ions. The shortened O:H nonbond is associated with energy and vibration frequency gain but the lengthened H-O bond is suffered frequency and energy loss.

The following generalizes $T_C(P)$ as a function of the segmental volume change $v_x = s_x d_x$ under pressure (x = H for the H-O bond and x = L for the O:H nonbond) with $s_x$ being the cross-sectional area of the segment,

$$\frac{T_C(P)}{T_C(P_0)} = 1 - \frac{E_x(P)}{E_{x0}} = 1 - \frac{\int_{V_0}^{V} p\, dv_x}{E_{x0}} = \begin{cases} > 0 & (d_L) \\ < 0 & (d_H) \\ \cong 0 & (\theta) \\ \cong \infty & (\theta) \end{cases}$$

(8)

The $E_{x0}$ is the reference of the segmental cohesive energy at the ambient conditions. The $d_L$, $d_H$, and $\theta$ are factors dominating the respective $T_C(P)$ profile.

The $T_C(P)$ profiles for the Liquid-Ic and the Ice VII-VIII phase transition epitomize the situation of negative slopes. With the known $d_x(P)$ relation, the following proves that $E_H$ relaxation dictates these $T_C(P)$. According to eq (5), the $T_C$ changes with $E_x$ but $x = H$ is as yet to be confirmed [57],

$$\frac{T_C(P)}{T_C(P_0)} = 1 - \frac{s_H \int_{P_0}^{P} p \frac{dd_H}{dp} dp}{E_{H0}} < 1.$$

(9)

Reproduction of the measured $T_C(P)$ for the VII–VIII phase transition [51, 52, 56], and the $T_m(P)$ for ice melting (-22 °C at 210 MPa; +6.5 °C at -95 MPa, see Figure 1b) [6, 74] requires that the integral must be positive, or the $d_x$ decreases when the pressure is applied. Only the $d_H$ in Eq. (1), meets this criterion. Therefore, the H-O bond dominates the $T_m$ and the $T_C$, as shown in Figure 1b and Figure 17. With the known $T_C(P)$ and the known $d_x(P)$ profiles one can determine the $E_{x0}(P_0)$, which specifies that the H-O bond elongation under compression results in the phase boundaries with $dT_C/dP < 0$.

Furthermore, matching both the $T_m(P)$ and the $T_C(P)$ profiles yields an $E_H$ value of 3.97 eV by taking the H atomic diameter of 0.106 nm as the H-O bond diameter [75]. This $E_H$ value agrees with the energy of 4.66 eV for dissociating the H-O bond of water molecules deposited on a $TiO_2$ substrate with less than a monolayer coverage, and 5.10 eV for dissociating water monomers in the gaseous phase [76]. Molecular undercoordination differentiates values of 5.10, 4.66 and 3.97 eV for the H-O bond in various coordination environments [77]. Reproduction of both the $T_C(P)$ for VII-VIII and $T_m(P)$ indicates that the $d_x(P)$ follows the same quantitative relationship given in Eq. (1) in these two situations.

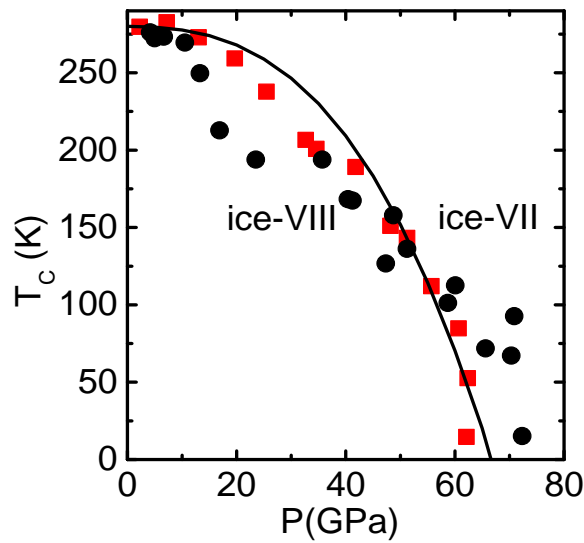

Figure 17 Theoretical reproduction of the measured $T_C(P)$ for the VII-VIII phase transition and the $T_m(P)$ (Figure 1b) confirms that the $E_H$ dictates the $T_C(P)$ with derivative of $E_H$ = 3.97 eV for bulk water and ice. (Reprinted with permission from [7, 10].)

7.3 O:H nonbond contraction dominance: $dT_C/dP > 0$

With the known $T_C(P)$ profiles of $dT_C/dP > 0$, one can clarify that such boundaries arise from O:H length depression and derive the $d_x(P)$ equation. A numerical fitting to the $T_C(P)$ profile for the Liquid-Vapor transition [69, 78], see Figure 18a, for instance, yields the following,

$$\frac{T_C(LnP)}{225.337} = 1 + 0.067757 \times \exp\left(\frac{LnP}{5.10507}\right) = 1 + A\exp\left(\frac{LnP}{B}\right) = 1 + AP^{\frac{1}{B}}$$

(10)

Equaling eqs (8) and (10),

$$AP^{\frac{1}{B}} = -D\int_{V_0}^{V} p\frac{dd_L}{dp} dp$$

(11)

yields the pressure dependence of the O:H length,

$$d_L(P) = -\frac{A}{(B-1)D}\exp\left(\frac{-LnP}{1.2436}\right) = Const \times P^{-0.80412}$$

(12)

Figure 18b plots the pressure trend of the O:H (approaches to O-O in vapor) length at the Liquid-Vapor phase boundary obtained by taking the $d_L$(0.1MPa, 333K) as unity of standard. Indeed, the slope of the $d_L(P)$ is negative. Likewise, one can obtain the pressure trend of the O:H distance iterating the same for a specific phase boundary of $dT_C/dP > 0$.

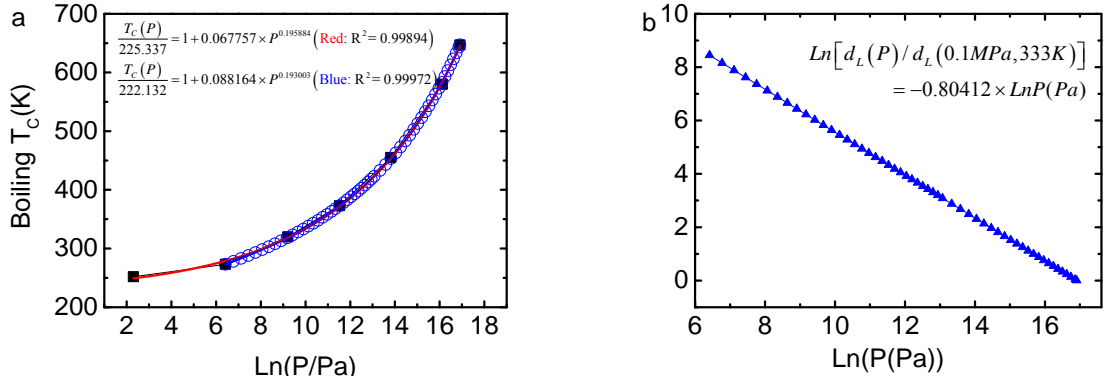

Figure 18 (a) Fitting the $T_C(P)$ for Liquid-Vapor phase transition (sourced from [69, 78] at different pressure ranges) with derivative of (b) the pressure trend of the $d_L(P)$ in the Liquid-Vapor phase boundary.

7.4 Phase boundaries of $dT_C/dP \cong 0 \text{ or } \infty$

The phase boundaries of $dT_C/dP \cong 0$ or $\infty$ means that the $T_C$ is energy independent. The O:H-O cooperative relaxation in length and energy contributes insignificantly to the phase transition at these boundaries but bond angle relaxation changing the geometry configuration of ice.

$T_C(P) \approx 70$ K at the Ic-XI phase boundary is within the IVth temperature regime where the specific heat $\eta_x \approx 0$ [9]. Neither the O:H nor the H-O bond is subject to length and energy change but the O:H-O containing angle become dominance in the structure relaxation [9].

The phase boundaries meet the criterion, $dT_C/dP \cong \infty$, or $T_C \approx \delta(P_C)$, at rather high pressures are insensitive to temperature, typically the boundaries of X-XI, X-(VIII,VIII). $\delta(P_C) = 1$ if $P = P_C$, else, $\delta(P_C) = 0$. The O:H and the H-O are identical in length of 0.11 Å in phase X [10]. Insignificant length relaxation happens to both segments at these boundaries but geometrical evolution.

7.5 Mechanical freezing of ambient water

When subjected to about 1 GPa compression, liquid water turns to ice-VI at room temperature [7]. Figure 19 shows the pressure-dependent Raman spectra of water at 25°C. During the Liquid-VI phase transition, the pressure suddenly drops from 1.35 to 0.86 GPa although the volume of the diamond compression cell containing the water sample shrinks continually [79]. At transition, a sudden blueshift occurs to both the $\omega_H$ and $\omega_L$. The sharp features indicate ice formation. The blueshift of the $\omega_x$ indicates that both the O:H and H-O undergo contraction according to Figure 19, which disobeys their correlation in eq (2).

Assuming the volume of the diamond cell remains unchanged, molecular separation will change approximately with $d_{OO} \propto P^{-1/3}$. Therefore, the $d_{OO}$ changes from its liquid phase to ice VI phase by $(1.35/0.86)^{-1/3} \approx 0.86$, which means that the O-O shrinks spontaneously by 14%. The density increases by 57% upon solidification. In the ice VI phase, the $\omega_x$ shifts cooperatively, following the trend of compressed ice at lower temperatures, shown in Figure 10. The $\omega_H$ undergoes a redshift but $\omega_L$ a blueshift.

The sudden drop in pressure and the blueshift of the $\omega_x$ upon freezing indicates that both the O:H and H-O deformation dominates the liquid-VI transition. The simultaneous O:H and H-O contraction suggests that the O-O Coulomb repulsion become weaker in the ice VI phase compared with liquid water.

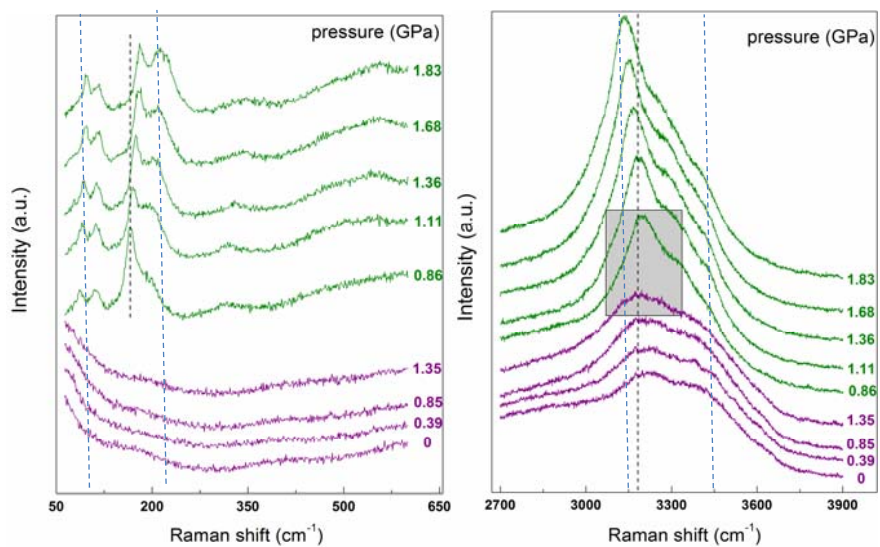

Figure 19. Raman spectra of the mechanico-freezing of ambient deionized water (25 °C) reveals a sudden drop in pressure from 1.35 to 0.86 GPa at freezing. The simultaneous blueshift of the $\omega_L$ and $\omega_H$ indicates that the O:H nonbond and the H-O bond undergo contraction. Hatched vertical lines denote the $\omega_x$ of the skin (75 and 3450 cm$^{-1}$), bulk (< 300 and 3200 cm$^{-1}$) and the $\omega_H$ for bulk ice (3150 cm$^{-1}$). The skin feature at 3450 cm$^{-1}$ persists throughout the applied pressures, which indicates that water contacts diamond hydrophobically. (Reprinted with permission from [79].)

7.6     X:H-O (X = N, F, Cl) bond relaxation

Asymmetrical, short-range O:H-O bond potentials are intrinsic to specimens containing F, O, and N element. The short-range interactions and Coulomb coupling is applicable to inter- and intramolecular interactions of these materials. For instance, Raman measurements have revealed coupled $\omega_L$ stiffening (110–290 cm$^{-1}$) and $\omega_H$ softening ($\approx$ 3000 cm$^{-1}$) in the O:H-N bonds in oxamide subjected to compression [44]. The pressure-trend of the Raman shifts of Oxamide (CO(NH$_2$)$_2$) [80] and Biurea (C$_2$H$_6$N$_4$O$_2$) [81] super molecules, shown in Figure 20, exactly emulate the trend of compressed water ice [82].

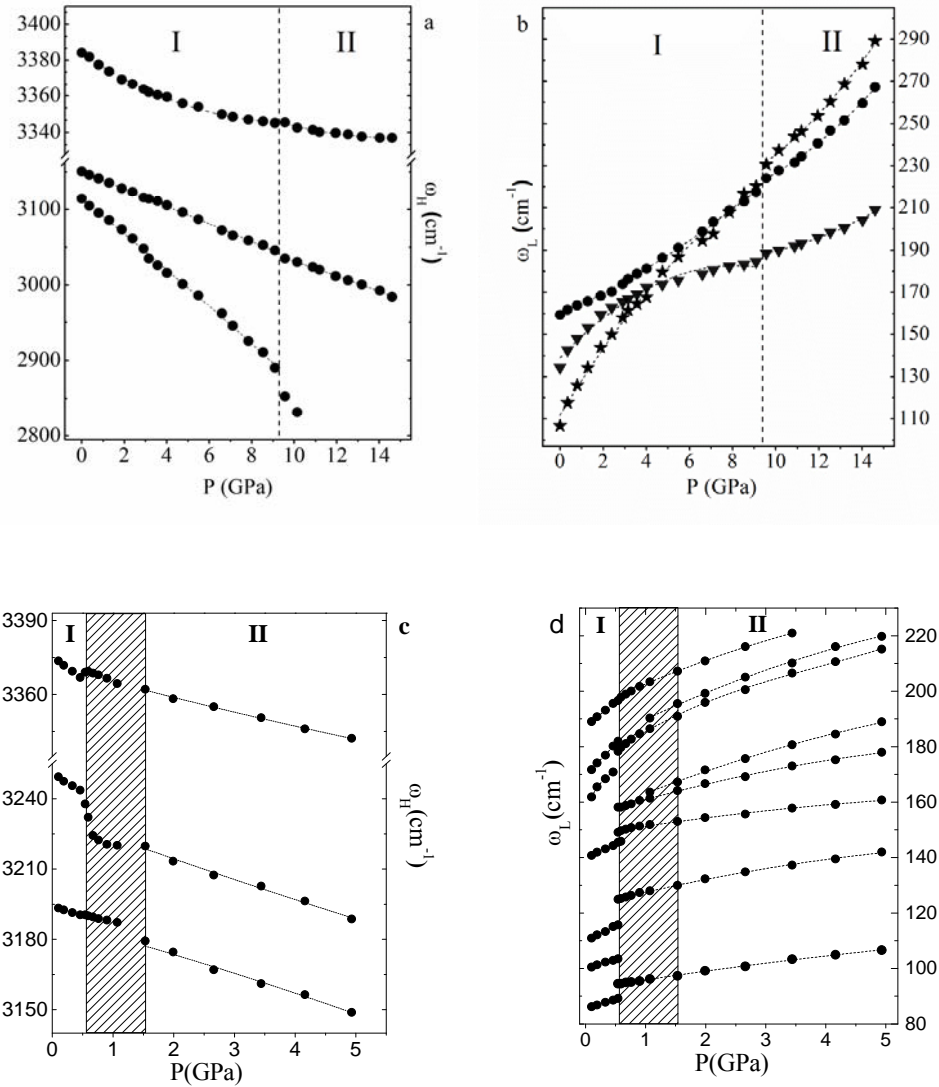

Figure 20. Compression stiffens the $\omega_L$ and softens the $\omega_H$ phonon of the O:H-N bond in (a, b) Oxamide (CO(NH$_2$)$_2$) and (c, d) Biurea (C$_2$H$_6$N$_4$O$_2$) molecular crystals at different phases I and II, indicating the presence of inter electron pair repulsion coupling the O:H-N bond. (Reprinted with permission from [44, 81].)

Compression at pressures greater than 150 GPa also softens the phonons of hydrogen crystal ($\approx$ 4000 cm$^{-1}$) at various temperatures [83]. Computations reveal that compression symmetrizes the intra- and inter-H$_2$ molecular distance [84]. These observations may indicate that short-range inter- and intramolecular interactions and the Coulomb coupling exist in hydrogen crystals.

Figure 21 shows the pressure dependence of the (a) $\omega_L$ spectra, (b) $\omega_L$ shift, (c) X:H-O length and, (d) $\omega_H$ for the F:H-O and Cl:H-O bond in $CuF_2(H_2O)_2$(3-chloropyridine)[85]. The change in line color denotes a new phase (or coexistence of phases). HP-I is the first high pressure phase, HP-II is the second high pressure phase, AFM is antiferromagnetic, and FM is ferromagnetic.

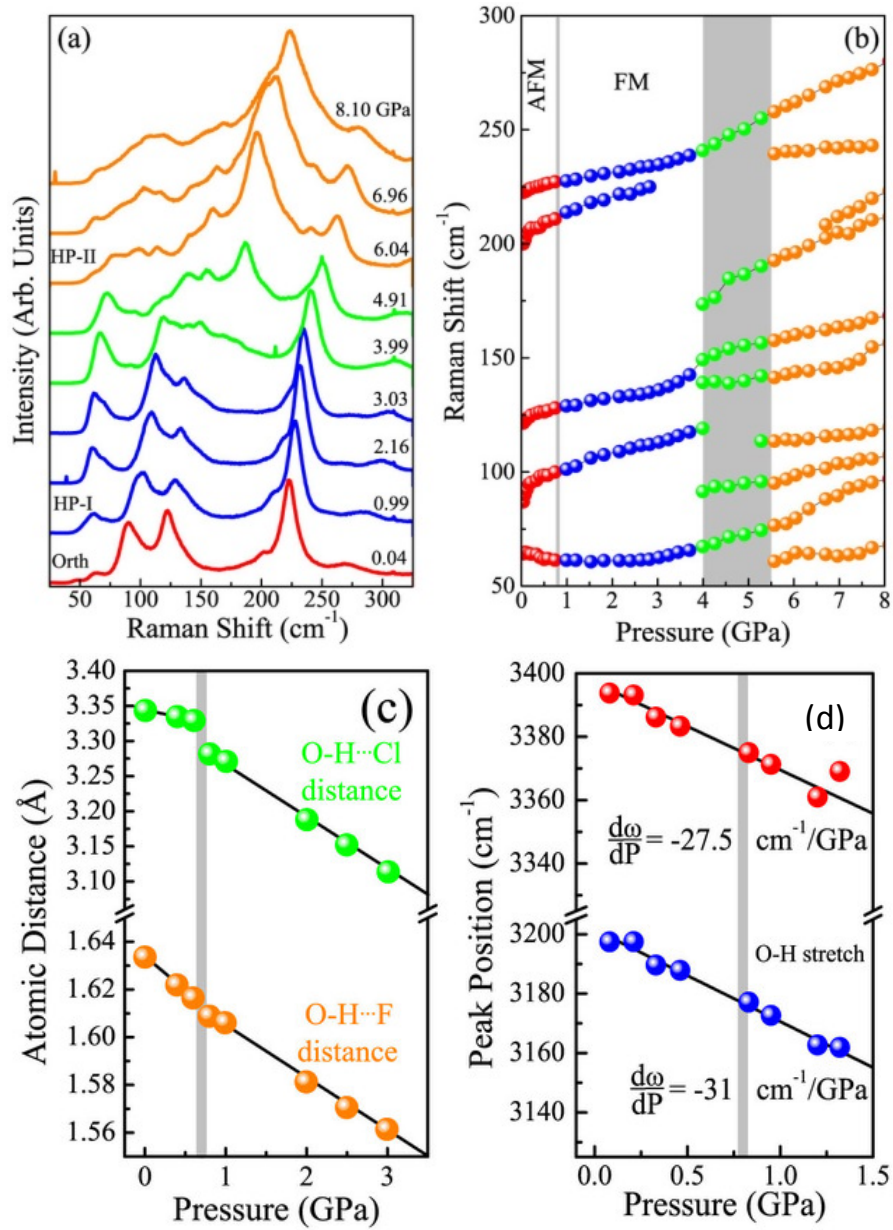

Figure 21. Pressure dependence of the (a) X:H stretching phonon $\omega_L$ spectra, (b) $\omega_L$ shift, (c) X:H-O length, and (d) H-O stretching phonon $\omega_H$ for $CuF_2(H_2O)_2$(3-chloropyridine). (Reprinted with permission

from [85].)

Likewise, all the X:H-O bonds share the same pressure trend of segmental length and stiffness relaxation. Compression shortens the X:H length and stiffens its stretching phonon $\omega_L$ but lengthens the H-O distance slightly and soften its stretching phonon $\omega_H$, resulting in the X:H-O contraction or density gain. The La:F [86] and Fe:S [87] stretching phonons at $\omega_L \leq 500$ cm$^{-1}$ also undergo compression stiffening.

Therefore, X:H-O bond exists in a wide range of materials — $H_2O$, $NH_3$, HF, $H_2$, oxides, nitrides and fluorides — because of the presence of short-range interactions. N, O and F create nonbonding lone pairs upon reacting with atoms of other less-electronegative elements. Based on the current notation of O:H-O bond cooperativity, it is expected that asymmetrical relaxation in length and stiffness of the O:H-O bond dictates the functionality of species with O:H-O bond-like involvement, including biomolecules, organic materials, H crystals, among others.

8       Summary

Regelation arises from the O:H-O bond memory and the phase boundary dispersivity:

1) Compression shortens and stiffens the O:H nonbond and lengthens and softens the H-O bond in all phases towards proton centralization, which lowers the compressibility of water ice and makes the O:H contribute positively, while the H-O contributes negatively to the lattice energy of compressed ice.
2) Compression closes up the separation between boundaries of the quasi-solid phase and depresses the $T_m$, resulting in ice regelation. Negative pressure does the opposite.
3) Unlike the bond in a 'normal' substance that gains energy with plastic deformation, compression raises the O:H-O bond to higher energy states monotonically. When the pressure is relieved, the O:H-O bond recovers completely its initial states with memory.
4) The O:H nonbond compression dictates the $T_C(P)$ phase boundaries with positive slopes like Liquid-Vapor transition - boiling;  the H-O bond elongation dictates the $T_C(P)$ phase boundaries with negative slopes like melting and VII-VIII transition; O:H-O length and energy conserve at boundaries of zero or infinite slopes like (XII,XIII)-X and IC-VI transition.